\documentclass[aps,prb,preprintnumbers,amsmath,amssymb,superscriptaddress,twocolumn,10pt]{revtex4-2}
\usepackage{txfonts}
\usepackage{graphicx}
\usepackage{dcolumn}
\usepackage{bm}
\usepackage{array}
\usepackage[dvipsnames]{xcolor}
\usepackage[%
    colorlinks=true,
    pdfborder={0 0 0},
    linkcolor=blue,
    citecolor=blue,
    urlcolor=blue
]{hyperref}  
\usepackage{chngpage}
\usepackage{appendix}
\usepackage[caption=false]{subfig}
\usepackage{booktabs}

\newcommand{\expectation}[1]{\langle#1\rangle}
\newcommand{\mbfk}{\mathbf{k}}
\newcommand{\mbfq}{\mathbf{q}}

\newcommand{\mbfd}{\mathbf{d}}
\newcommand{\mbfM}{\mathbf{M}}

\newcommand{\be}{\begin{equation}}
\newcommand{\ee}{\end{equation}}
\newcommand{\bea}{\begin{eqnarray}}
\newcommand{\eea}{\end{eqnarray}}

\begin{document}

\title{Symmetry-Enforced Pair-Density Wave and Chiral Interband Superconductivity in Strongly Correlated Kagome Systems}

\author{Alex Friedlan}
\affiliation{Department of Physics, University of Toronto, 60 St. George Street, Toronto, Ontario, Canada, M5S 1A7}
\author{Hae-Young Kee}
\email{hy.kee@utoronto.ca}
\affiliation{Department of Physics, University of Toronto, 60 St. George Street, Toronto, Ontario, Canada, M5S 1A7}
\affiliation{Canadian Institute for Advanced Research, CIFAR Program in Quantum Materials, Toronto, Ontario, Canada, M5G 1M1}

\begin{abstract}
The pair-density wave (PDW) state, characterized by Cooper pairing at finite momentum, is a long-sought superconducting phase whose possible realization in Kagome metals is particularly intriguing in the strongly correlated regime. We investigate superconductivity in the extended $t$-$J$ model on the Kagome lattice and show that the symmetry-enforced sublattice structure of the Bloch wavefunctions gives rise to a rich landscape of unconventional pairing states. When the chemical potential is tuned to a sublattice-pure ($p$-type) van Hove singularity (vHS), a PDW state inevitably emerges. Near the $m'$-type vHS, which features opposite mirror eigenvalues to the conventional $m$-type vHS, intraband chiral, uniform, and nematic pairing states compete. When further-neighbor hoppings drive the $p$- and $m'$-type vHSs towards near degeneracy, phase frustration in the interband pairing channel stabilizes a chiral interband state. Our results reveal the previously overlooked $m'$-type vHS as a distinct route to unconventional superconductivity rooted in electronic correlations and mirror-symmetry-constrained Bloch wavefunctions.
\end{abstract}
\date{\today}
\maketitle

{\it Introduction.}---A pair-density wave (PDW) is a superconducting state in which a Cooper-pair condensate carries a finite center-of-mass (COM) momentum, exhibiting modulation in real space \cite{Agterberg2020PDW,WangPDWReview2026}. Unlike the well-known FFLO state \cite{FF1964,LO1964,Matsuda2007FFLOreview}, where the finite-COM-momentum pairing arises from the Zeeman splitting associated with the spin imbalance induced by an external magnetic field, a PDW is considered a strong-interaction-driven state. Although PDWs have been proposed in several candidate materials, including high-$T_c$ cuprates \cite{Berg2009,FradkinColloquiumPDW2015,LeePRXPDW2014,Hamidian2016,Chen2004,Himeda2002,Chen2022NematicPDW}, iron-based superconductors \cite{Paglione2010,Zhao2023pnictide,Liu2023PDW}, and heavy-fermion superconductors \cite{Gu2023PDW,Aishwarya2023PDW}, there is still no universally accepted smoking-gun signal.

Among several candidates, the proposal of a PDW in the Kagome superconductors \cite{ChenH2021,DengPDW2024,WuPDWKagome2023,Yan2026KagomePDWPNAS,Song2026ChiralPDW,YanXY2024ChiralPDW,Peng2026GL} is motivated by the interplay between distinctive features of the Kagome electronic structure and the experimental phenomenology observed in AV$_3$Sb$_5$ (A = K, Rb, Cs) \cite{Ortiz2019,WangReview2023,DiSante2026}. However, these materials host a charge-density wave (CDW) above the superconducting transition temperature \cite{Ortiz2020,Zhao2021,Yuarxiv2021,Mielke2022,Jiang2021NatureMat,Xing2024,XuY2022,Xiang2021,Li2022,Wu2022PRB,Nie2022,ZhengL2022}, which generically induces a PDW at half the CDW wavevector via the coupling between the two order parameters \cite{Agterberg2008,Berg20094e,Du2020}. As a result, a genuine PDW independent of the CDW has recently been explored in several studies \cite{YaoM2025,Lamponen2025, GenuinePDW, ZhangY2025}. Unconventional superconductivity in the Kagome metals more broadly has also generated significant interest \cite{Jiang2021,Neupert2022,Wilson2024,Roppongi2023,Guguchia2023,Fukushima2024,Daniel2025,Nana2021,Deng2024,ZhouS2022,Wu2021,LiuY2024_Cr,WangZ2025_Cr}.

Here we investigate possible superconducting states on the Kagome lattice in the strongly interacting regime. Starting from the Hubbard model in the large-$U$ limit, a nearest-neighbor (NN) Heisenberg exchange interaction is derived. Owing to geometric frustration, magnetic ordering is suppressed, while the exchange interaction $J$ provides an attractive channel for spin-singlet pairing. Within the resulting $t\text{-}J$ model, the symmetry of the lattice plays a central role in determining the nature of the pairing \cite{Kiesel2012}. In particular, the Kagome lattice features van Hove singularities (vHS) at the $M$ points with distinct mirror symmetries. When the vHS is sublattice-pure ($p$-type), conventional zero-COM-momentum pairing is forbidden and a PDW state emerges, whereas proximity to the sublattice-mixed ($m$-type) vHS reestablishes zero-COM-momentum pairing.

While previous studies of superconductivity have focused on these two conventional vHSs \cite{YuSL2012, Romer2022, Schwemmer2024,Holbaek2023PRB,LiuJ2024KagomeSymmetry,Wang2013,Wen2022}, the Kagome lattice also hosts a third vHS at $M$, denoted $m'$-type, which evolves out of the canonical Kagome flat band in the presence of further-neighbor hoppings. Its distinct mirror symmetries allow for unconventional pairing states, including chiral and nematic superconductivity. Moreover, tuning further-neighbor hoppings can bring the $p$- and $m'$-type vHSs into close proximity, stabilizing a chiral interband pairing state. We begin below with the Hubbard model and present the phase diagram obtained by self-consistent mean-field (MF) theory.

\begin{figure}
    \centering
    \includegraphics[width=\linewidth]{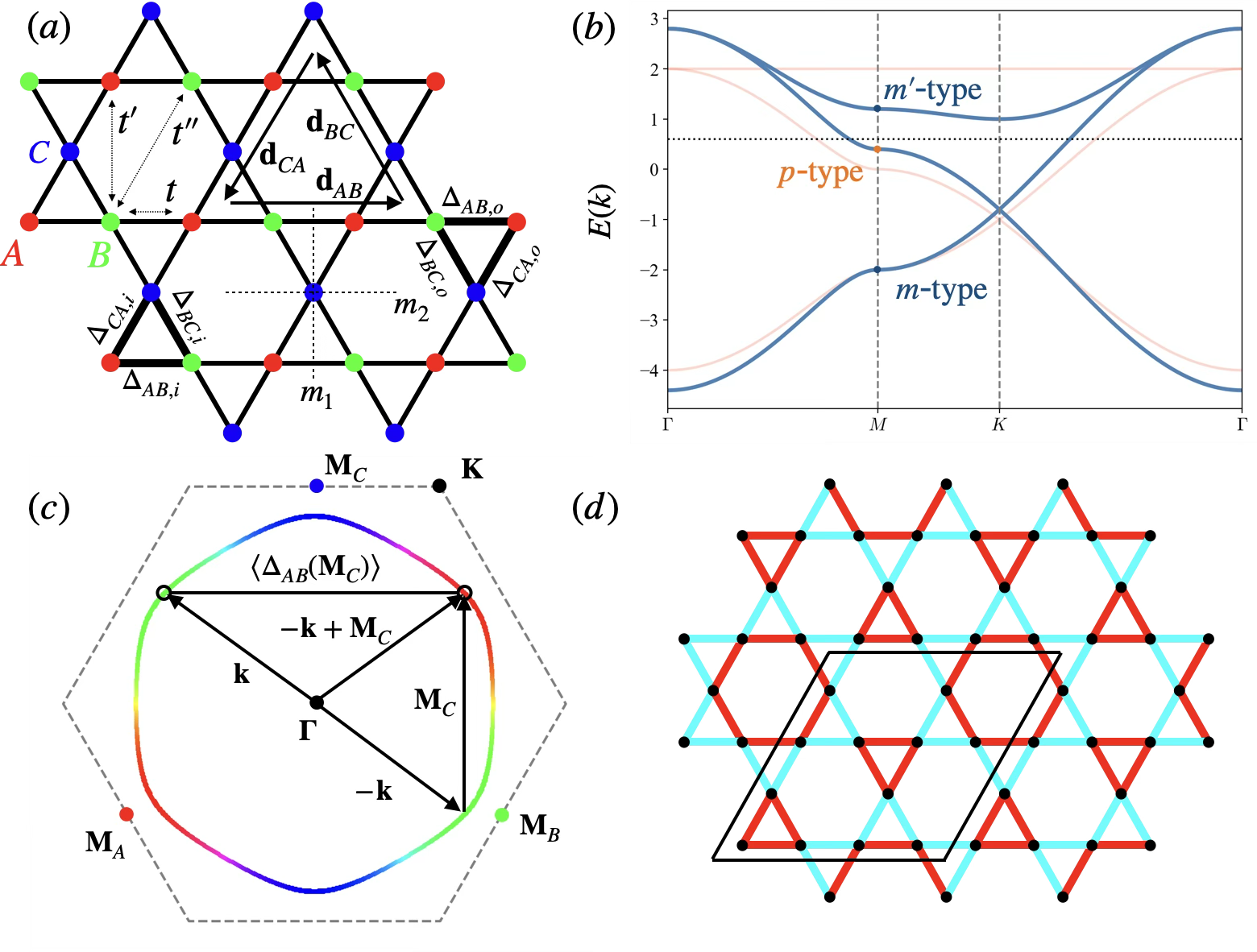}
    \caption{(a) Kagome lattice with NN, second-NN, and third-NN hoppings $t$, $t'$, and $t''$, respectively, and two mirror operators ${\hat m}_1$ and ${\hat m}_2$. (b) Single-particle dispersion for $t=-1$, $t'= -t'' = -0.2$. Ideal Kagome bands (orange) with $t'=t''=0$ are also shown for comparison. (c) Sublattice-resolved Fermi surface near the $p$-type vHS for $\mu=0.6$ ($n\approx 0.45$). Pairing $\langle \Delta_{AB} \rangle$ with COM momentum $\mbfq=\mbfM_C$ connects portions of the FS with $A$ and $B$ sublattice character. (d) The $2a\times2a$ PDW ($\mbfq=\mbfM$) in real space. Red/blue bonds denote positive/negative pairing amplitudes.}
    \label{fig1}
\end{figure}

{\it Model Hamiltonian.}---We consider the single-orbital Hubbard model on the Kagome lattice.
Taking the large-$U$ limit, and applying the Schrieffer-Wolff transformation \cite{MacDonald1988,Fazekas,tJthenandnow}, the low-energy effective Hamiltonian is given by the standard $t\text{-}J$ model [see Supplemental Material (SM) for details]:
\begin{equation}
    H_{\rm eff}= \sum_{ij} \sum_\sigma (t_{ij}-\mu\delta_{ij})  c_{i\sigma}^\dagger c_{j\sigma} + J \sum_{\langle ij\rangle} \left( \mathbf{S}_{i}\cdot \mathbf{S}_{j}-\frac{1}{4}n_i  n_j \right),
    \label{H_eff}
\end{equation}
where $c^\dagger_{i\sigma}$ creates an electron with spin $\sigma$ at site $i$, $t_{ij}$ is the hopping integral between sites $i$ and $j$, and $\mu$ is the chemical potential. For NN bonds $\langle..\rangle$, $t_{ij} \equiv t$ and $J = 4t^2/U$. Second- and third-NN hoppings are denoted $t'$ and $t''$, respectively, as illustrated in Fig.~\ref{fig1}(a). With their inclusion, the canonical Kagome flat band acquires a dispersion and a vHS at $M$ [see Fig.~\ref{fig1}(b)]. While $t'$ and $t''$ give rise to $J'= 4t'^2/U$ and $J''=4t''^2/U$, we have verified that including these further-neighbor exchange interactions does not significantly alter our conclusions (see SM); all other further-neighbor hoppings are neglected.

{\it Mirror symmetries and Bloch wavefunctions.}---To understand the emergence of the PDW and chiral interband pairing states, we first review the Bloch wavefunctions, whose sublattice structure is constrained by mirror symmetries~\cite{HeqiuPRL}, also known as the sublattice interference mechanism \cite{Kiesel2012}. As shown in Fig.~\ref{fig1}(a), the Kagome lattice possesses two mirror symmetries at each site, represented by the mirror operators ${\hat m}_1$ and ${\hat m}_2$. Consequently, Bloch wavefunctions at high-symmetry momenta that preserve these symmetries, such as the $M$ points, must be eigenstates of the corresponding mirror operators. The allowed forms of the wavefunctions for the three bands at the $M_\alpha$ points ($\alpha = A, B, C$) are summarized in Table~\ref{tabel}. Here, the three entries in parentheses denote the orbital weight on sublattices $A$, $B$, and $C$, respectively. For example, a zero in the first entry indicates no contribution from the Wannier orbital on sublattice $A$. We refer to states with weight localized on a single sublattice at a given $M$ point as $p$-type, while states with mixed sublattice character are denoted as $m$- and $m'$-type, as shown in Table~\ref{tabel}. As illustrated in Fig.~\ref{fig1}(c), close to a $p$-type vHS the sublattice-resolved FS shows clear contributions from the $A$ (red), $B$ (green), and $C$ (blue) sublattices at $M_A$, $M_B$, and $M_C$, respectively. Away from these mirror-symmetric $M$ points, the sublattice weights are mixed, as shown by the mixed-color regions.

\begin{table}
\centering
\renewcommand{\arraystretch}{1.4}
\setlength{\tabcolsep}{12pt}
\begin{tabular}{l ccc}
\toprule
& $M_A$ & $M_B$ & $M_C$ \\
\midrule
$m'$-type & $(0,\, b,\, {-b})$ & $({-b},\, 0,\, b)$ & $(b,\, {-b},\, 0)$ \\
$p$-type  & $(a,\, 0,\, 0)$    & $(0,\, a,\, 0)$    & $(0,\, 0,\, a)$   \\
$m$-type  & $(0,\, b,\, b)$    & $(b,\, 0,\, b)$    & $(b,\, b,\, 0)$   \\
\bottomrule
\end{tabular}
\caption{Orbital weight on sublattices $(A,B,C)$ at the $M_A$, $M_B$, and $M_C$ points for each of the three bands, where $a=1$ and $b=1/\sqrt{2}$. Although orbital hybridization in real materials may lead to quantitative changes in these values, the qualitative conclusions of our work remain unchanged.}
\label{tabel}
\end{table}

{\it Correlated metal and mean-field theory.}---In the absence of symmetry breaking, the system remains a correlated metal. Among several possible broken-symmetry states, a CDW order generally requires a repulsive NN interaction of the form $V n_i n_j$ \cite{Zhan2026,Kiesel2013,Denner2022,HeqiuPRB,HeqiuPRL}. In the large-$U$ limit, however, the superexchange interaction generates a term $-\frac{J}{4} n_i n_j$, as shown in Eq.~(\ref{H_eff}), which partially offsets the effect of the NN repulsion $V$, thereby suppressing charge-ordering tendencies. Meanwhile, the Kagome geometry frustrates magnetic ordering \cite{HeLW2025_KagomeQSLReview}, motivating us to focus on superconducting (SC) instabilities.

The SC order parameter is defined on NN bonds connecting different sublattices denoted by $\alpha$ and $\beta$. For each sublattice pair, there is a bond belonging to an upper Kagome triangle ($\triangle$), which lies inside the unit cell and is denoted by $\Delta_{\beta\alpha,i}$, and a bond belonging to a lower triangle ($\triangledown$), which connects neighboring unit cells and is denoted by $\Delta_{\beta\alpha,o}$; see Fig.~\ref{fig1}(a). There are therefore six independent order parameters, and they are given in operator form by
\begin{align}
   & {\hat \Delta}_{\beta\alpha,i}(\mbfq)= \frac{J}{N} \sum_{\mbfk} \frac{1}{\sqrt{2}}\left( c_{\beta,\mbfq-\mbfk\uparrow} c_{\alpha,\mbfk\downarrow} - c_{\beta,\mbfq-\mbfk\downarrow}c_{\alpha,\mbfk\uparrow} \right), \nonumber\\
  &  {\hat \Delta}_{\beta\alpha,o}(\mbfq)= \frac{J}{N} \sum_{\mbfk}  \frac{1}{\sqrt{2}}\left( c_{\beta,\mbfq-\mbfk\uparrow} c_{\alpha,\mbfk\downarrow} - c_{\beta,\mbfq-\mbfk\downarrow}c_{\alpha,\mbfk\uparrow}\right)e^{-i \mbfk \cdot {\bf d}_{\alpha\beta}},
\end{align}
where $\mbfd_{\alpha\beta}$ is the distance between NN unit cells along the $\alpha\beta$-bond direction ($\alpha,\beta\equiv A, B, C$), as shown in Fig.~\ref{fig1}(a). Then, the mean-field Hamiltonian is found as
\begin{align}
   & H_{\rm MF} = \sum_{\sigma,\alpha\neq\beta,\mbfk} t_{\mbfk}^{\alpha\beta} c^\dagger_{\alpha,\mbfk\sigma} c_{\beta,\mbfk\sigma} 
    +\frac{N}{J}\sum_{\alpha\neq\beta,\mbfq} \Big(|\Delta_{\beta\alpha,i}(\mbfq)|^2 + |\Delta_{\beta\alpha,o}(\mbfq)|^2\Big)\nonumber\\
&+\sum_{\alpha\neq\beta,\mbfk,\mbfq} \frac{1}{\sqrt{2}}
\left( \Delta_{\beta\alpha,i}^*(\mbfq) + \Delta_{\beta\alpha,o}^*(\mbfq)e^{-i\mbfk\cdot\mbfd_{\alpha\beta}} \right)\left(  c_{\beta,\mbfq-\mbfk\uparrow} c_{\alpha,\mbfk\downarrow} - \downarrow  \uparrow
   \right)  + \text{h.c.},
\end{align}
where the hopping $t_{\mbfk}^{\alpha\beta}$ includes the $(t,t',t'',\mu)$ terms and is presented in the SM.

 \begin{figure*}
     \centering
     \includegraphics[width=\linewidth]{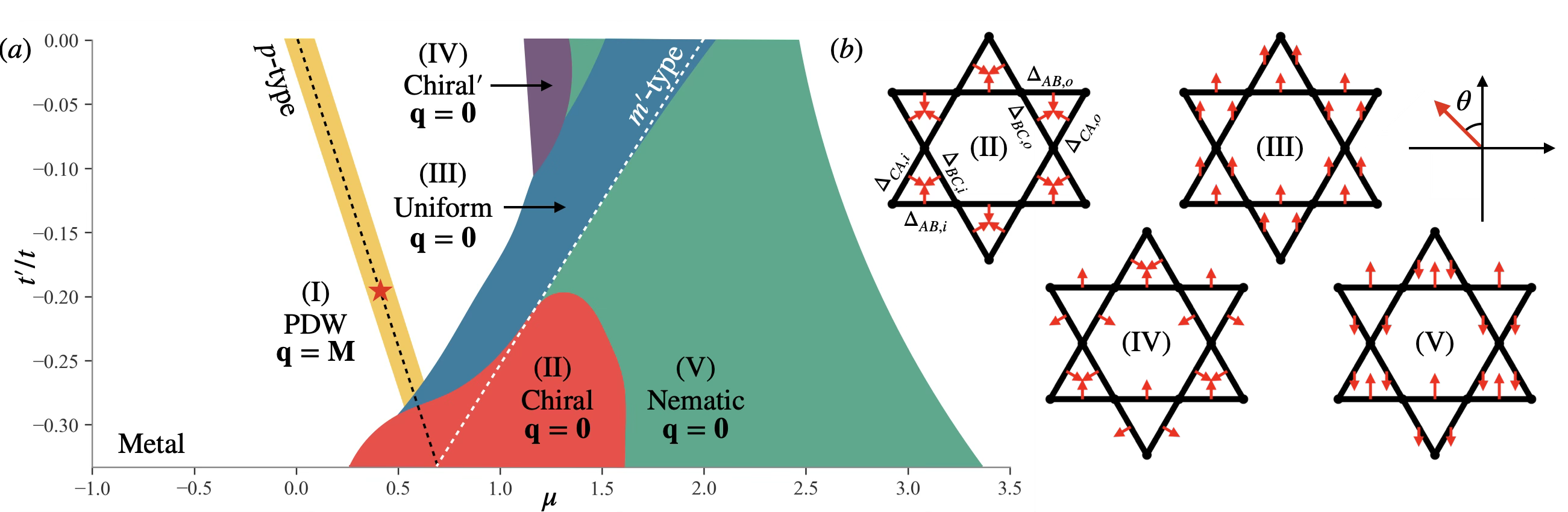}
     \caption{(a) Phase diagram in the $\mu$-$t'$ plane ($t'' =-t'$) at $U = 2|t|$ and zero temperature. The black and white dashed lines indicate the locations of the $p$- and $m'$-type van Hove singularities at the $M$ point, respectively. The red star corresponds to the point examined in Fig.~\ref{fig1}. (b) Real-space representations of the phases. The amplitude and phase $\theta$ (defined in the upper-right corner) of the pairing on each bond are represented by arrows. The order parameter is characterized by the tuple $(\Delta_{AB,i},\Delta_{BC,i},\Delta_{CA,i},\Delta_{AB,o},\Delta_{BC,o},\Delta_{CA,o})$: (I) PDW, see Fig.~\ref{fig1}(d); (II) Chiral order $(E_1)$: $(1,\omega,\omega^2,-1,-\omega,-\omega^2)$, where $\omega=e^{i2\pi/3}$; (III) Chiral' order $(E_2)$: $(1,\omega,\omega^2,1,\omega,\omega^2)$; (IV) Uniform order $(A_1)$: $(1,1,1,1,1,1)$; (V) Nematic order: $(1,-\eta,-\eta,1,-\eta,-\eta)$. The momentum-space pairing structures for all phases are shown in the SM.}
     \label{fig2}
 \end{figure*}

{\it Pair-density wave near the p-type vHS.}---When the chemical potential is tuned to the vicinity of a $p$-type vHS, a mirror-symmetry-enforced PDW is stabilized, characterized by finite $\Delta_{\alpha\beta,i}(\mbfM_\gamma) = -\Delta_{\alpha\beta,o}( \mbfM_\gamma)\equiv \Delta$, where $\Delta$ is real and $\alpha\neq\beta\neq\gamma$. Pairing with COM momentum $\mbfM_\gamma$ is inevitable because the sublattice structure of the FS implies that the opposite-momentum partners $\pm\mathbf{k}$ carry weight on the same sublattice, whereas the interaction favors pairing between different sublattices; see Fig.~\ref{fig1}(c). As a result, Cooper pairs are formed between quasiparticles near two distinct $M$ points, for example $M_\alpha$ and $M_\beta$, giving rise to a PDW with COM momentum $\mbfM_\gamma$. The real-space structure of the PDW with its enlarged unit cell is shown in Fig.~\ref{fig1}(d).

The MF phase diagram is obtained as a function of $\mu$ and $t'$ (with $t'' = -t'$ for simplicity) at $U=2|t|$ and is shown in Fig.~\ref{fig2}(a). The chosen parameter range covers only the upper two vHSs ($p$- and $m'$-types), while the lower $m$-type vHS occurs at smaller $\mu$. The PDW (yellow region) persists near the $p$-type vHS as we tune $t'$. As $\mu$ is tuned away from the $p$-type vHS, the PDW is suppressed, but several $\mbfq =0$ pairing channels emerge in the vicinity of the $m'$-type band (purple, blue, and green regions). We will discuss these phases in more detail later. First, we focus on the emergence of a chiral interband phase (indicated by the red region), which develops when the $p$- and $m'$-type vHSs are brought into close proximity by tuning $t'$. If $U$ is increased, the PDW stripe becomes very narrow, eventually disappearing for $U>4|t|$. Meanwhile, the interband chiral region contracts towards the $p$-$m'$ degeneracy point. The boundaries of the other phases contract as well, but to a lesser degree as they are primarily of intraband nature (see SM).

{\it Chiral interband pairing.}---This phase is characterized by a finite $\Delta_{\alpha\beta,i}(\mbfq=0)=-\Delta_{\alpha\beta,o}(\mbfq=0)$. Suppressing the momentum dependence (and henceforth focusing on $\mbfq=0$ pairing), each SC phase can be characterized by the tuple $(\Delta_{AB,i},\Delta_{BC,i},\Delta_{CA,i},\Delta_{AB,o},\Delta_{BC,o},\Delta_{CA,o})$. The Chiral phase then takes the form $\Delta(1,\omega,\omega^2,-1,-\omega,-\omega^2)$, where $\omega=e^{i2\pi/3}$; see Fig.~\ref{fig2}(b). As we will show, this phase arises primarily from interband pairing. Since $\mbfq=0$, this is not a PDW in the traditional sense, but instead features an internal structure in which the pairing amplitude changes sign between the bonds within the unit cell. While Ref.~\cite{Schwemmer2024} reported the possibility of pairing modulation within a unit cell, its microscopic origin and band-basis character have not yet been elucidated.

To understand the emergence of this Chiral phase, we examine the MF Hamiltonian near an $M$ point. In the basis $\Psi_\mbfk^\dagger = \left\{c_{A,\mbfk\uparrow}^\dagger,c_{B,\mbfk\uparrow}^\dagger,c_{C,\mbfk\uparrow}^\dagger,-c_{A,-\mbfk\downarrow},-c_{B,-\mbfk\downarrow},-c_{C,-\mbfk\downarrow}\right\}$, the $3\times 3$ off-diagonal pairing matrix is  given by
\begin{equation}
    \mathcal{H}_{\rm int}(\mbfk)=
    \begin{pmatrix}
        0 & \Delta_{AB,\mbfk} & \Delta_{CA,-\mbfk}  \\
        \Delta_{AB,-\mbfk} & 0 & \Delta_{BC,\mbfk} \\
        \Delta_{CA,\mbfk} & \Delta_{BC,-\mbfk} & 0 
    \end{pmatrix},
\end{equation}
where
\begin{equation}\label{momentumpairing}
\Delta_{\beta\alpha,\mbfk}\equiv \frac{1}{\sqrt{2}}\left(\expectation{{\hat \Delta}_{\beta\alpha,i}} + e^{i\mbfk\cdot\mbfd_{\alpha\beta}} \expectation{{\hat \Delta}_{\beta\alpha,o}}\right).
\end{equation}
Consider the pairing matrix at a particular $\mbfM_\alpha$, e.g. at $\mbfk = \mathbf{M}_C$, where $\mbfM_C \cdot {\bf d}_{AB} =0$ and $\mbfM_C \cdot {\bf d}_{BC} = \mbfM_C \cdot {\bf d}_{CA} = \pi$, fixing the phase relation between the two bond order parameters in Eq. (\ref{momentumpairing}). Applying the unitary transformation $U(\mbfk)={\rm{diag}} (U_{\mbfk},U_{\mbfk})$ at $\mbfk=\mbfM_C$ (see the SM for details), we obtain the pairing matrix in the band basis:
\begin{align}\label{gapmatrixC}
    & U^\dagger(\mathbf{M}_C) \mathcal{H}_{\rm int}(\mbfM_C)U(\mathbf{M}_C) \nonumber\\ 
    & = \begin{pmatrix}
        0 & \frac{1}{\sqrt2}(\Delta_{CA}^--\Delta_{BC}^-) & \frac{1}{\sqrt2}(\Delta_{CA}^-+\Delta_{BC}^-) \\
        \frac{1}{\sqrt2}(\Delta_{CA}^--\Delta_{BC}^-) & -\Delta_{AB}^+ & 0 \\
        \frac{1}{\sqrt2}(\Delta_{CA}^-+\Delta_{BC}^-) & 0 & \Delta_{AB}^+
    \end{pmatrix},
\end{align}
where the bands are ordered ($p,m',m$) and $\Delta_{\alpha\beta}^\pm\equiv \frac{1}{\sqrt{2}} \left(\expectation{{\hat \Delta}_{\beta\alpha,i}} \pm \expectation{{\hat \Delta}_{\beta\alpha,o}}\right)$. The gap matrix exhibits both intraband pairing $(\sim \Delta_{\alpha\beta}^+)$ and interband pairing $(\sim \Delta_{\alpha\beta}^-)$. The intraband gap function vanishes for the $p$-type band at $M$ due to sublattice interference. Therefore, to efficiently gap out the FS, the system resorts to {\it interband} pairing, which only becomes favorable when the $p$- and $m'$-type vHSs are brought into close proximity by tuning $t'$. To maximize the interband pairing amplitude $\frac{1}{\sqrt2}(\Delta_{CA}^- - \Delta_{BC}^-)$, the order parameters should satisfy $\Delta_{CA,i} = -\Delta_{CA,o}$ and $\Delta_{BC,i} = -\Delta_{BC,o}$, with $\Delta_{CA}^-$ and $\Delta_{BC}^-$ taking opposite signs. Note that interband pairing competes with intraband pairing, as maximizing $\Delta_{\alpha\beta}^-$ comes at the cost of minimizing $\Delta_{\alpha\beta}^+$ and vice versa.

The pairing matrices in the band basis at $M_A$ and $M_B$ can be obtained from Eq. (\ref{gapmatrixC}) by threefold rotation. To maximize interband pairing at these points as well, we see that the phases of the three components $\Delta_{AB}^-$, $\Delta_{BC}^-$, and $\Delta_{CA}^-$ prefer to differ by $\pi$. However, this constraint cannot be satisfied at all three $M$ points simultaneously, leading to frustration among $\Delta_{AB}^-$, $\Delta_{BC}^-$, and $\Delta_{CA}^-$. To relieve this frustration, the system adopts a configuration in which the relative phases differ by $2\pi/3$, resulting in a chiral $\mbfq=0$ state. The frustration arises from the relative sign between the order parameters in the interband matrix element, which in turn originates from the relative sign between orbital coefficients of the $m'$-type vHS (see Table~\ref{tabel}). This phase transforms according to the $E_1$ irreducible representation (irrep.) of $D_6$ \cite{Holbaek2023PRB}.

{\it Additional zero-COM-momentum pairings.}---Away from the $p$-type vHS but close to the $m'$-type vHS, several $\mbfq=0$ states emerge depending on the value of $t'$. These phases are all characterized by $\Delta_{\alpha\beta,i}(\mbfq=0)=\Delta_{\alpha\beta,o}(\mbfq=0)$, in contrast to the PDW and Chiral states discussed so far, where the two bond order parameters have opposite signs. In addition to the conventional uniform pairing with $s$-wave symmetry [blue region in Fig.~\ref{fig2}(a)], there are two other $\mbfq=0$ pairing states characterized by different phases among $\Delta_{AB}$, $\Delta_{BC}$, and $\Delta_{CA}$.

For one such phase, two of the pairing order parameters adopt a relative phase of $\pi$, while the third remains at zero phase, e.g. $\Delta(1,-\eta,-\eta,1,-\eta,-\eta)$ with $\eta<1$. Since this configuration breaks $C_3$ symmetry, we refer to it as the Nematic pairing state (green region), which extends over a large range of $t'$. For small $t'$, there is also a $C_3$-symmetric chiral state, denoted Chiral$'$ (purple region), in which the phases differ by $2\pi/3$, i.e. $ \Delta(1,\omega,\omega^2,1,\omega,\omega^2)$, where $\omega=e^{i2\pi/3}$. This phase is distinct from the Chiral phase discussed above, in that the pairing amplitude has the same sign on both bond types within the unit cell. As a consequence, the pairing is primarily of {\it intraband} nature, and transforms according to the $E_2$ irrep. of $D_6$ \cite{Holbaek2023PRB,LiuJ2024KagomeSymmetry} and breaks time-reversal symmetry.

In the regime near $t'=-t''=0$ where Chiral' develops, the Chiral', Uniform, and Nematic phases are nearly degenerate. Moreover, the free energy of the ordered state is only marginally lower than the free energy of the normal state due to the low DOS at the Fermi level. Although the Chiral' phase is consistently found to be a free-energy minimum as the $k$-mesh density is increased, we expect that going beyond MF will select a particular phase configuration, which is left for future work. At larger $U$, Chiral' moves to a different location in the phase diagram (see SM).

To understand the competition among these $\mbfq=0$ pairings near $t'=0$, we consider a Ginzburg-Landau (GL) free energy $F$. Taking advantage of the fact that the Uniform, Nematic, and Chiral' phases are all characterized by $\Delta_{\alpha\beta,i}=\Delta_{\alpha\beta,o}\equiv \Delta_{\alpha\beta}$, the quadratic contribution to $F$ takes the simple form
\begin{align}
    F_2 & = \sum_n a_1 |\Delta_{n}|^2 + \sum_{n\neq m} a_2\Delta_{n} \Delta_{m}^*,
\end{align}
where, for brevity, we label the three bonds by $n=1,2,3$, corresponding to $(AB,BC,CA) \leftrightarrow (1,2,3)$. The sign of $a_2$ selects between the Uniform and Chiral$'$ phases through the phase-sensitive factor $\sum_{n\neq m} \cos(\phi_n-\phi_m)$.  For $a_2>0$, the Chiral$'$ phase is favored, with $(\Delta_1,\Delta_2,\Delta_3)=\Delta(1,\omega,\omega^2)$, whereas for $a_2<0$, the Uniform phase $(\Delta_1,\Delta_2,\Delta_3)=\Delta(1,1,1)$ is selected. To describe the transition to the Nematic state, it is necessary to include the quartic contribution $F_4$:
\begin{align}
 & \sum_n b_1 |\Delta_{n}|^4 + \sum_{n\neq m}\left( b_2|\Delta_{n}|^2 |\Delta_{m}|^2 + b_3 \Delta_{n}^2 (\Delta_{m}^*)^2+ b_{4}|\Delta_{n}|^2\Delta_{n} \Delta_{m}^*\right)  \nonumber \\
    & +  \sum_{n\neq m \neq l} \left(b_5 \Delta_{n}^2 \Delta_{m}^*\Delta_{l}^* + b_6 |\Delta_{n}|^2 \Delta_{m} \Delta_{l}^* + \text{c.c.} \right).
\end{align}
The Nematic state can be understood as a balance between the $b_4$ and $b_5$ terms. The $b_4$ term contains the same phase-sensitive structure as the $a_2$ term and therefore favors the Chiral' phase for $b_4>0$ and the Uniform phase for $b_4<0$, while also allowing for nematicity among the amplitudes $|\Delta_n|$. Meanwhile, the $b_5$ term contains the factor $\sum_{n\neq m \neq l} \cos(2\phi_n - \phi_m-\phi_l)$, which favors the Chiral$'$ phase for $b_5<0$ and a nematic/chiral configuration $\Delta(1,\omega,\omega)$ for $b_5>0$. The competition between $b_4<0$ and $b_5>0$ stabilizes the Nematic phase $\Delta(1,-\eta,-\eta)$. A sextic term $\sum_n c_1 |\Delta_{n}|^6$ should also be included to bound the free energy (see SM for more details).

{\it Summary and Discussion.}---In summary, we have studied the extended $t$-$J$ model on the Kagome lattice and find that a PDW state emerges near the $p$-type vHS. The PDW arises as an unavoidable consequence of the sublattice structure at the $M$ point enforced by mirror symmetries, which constrains the pairing channels available to the low-energy states. Upon tuning the chemical potential such that the Fermi surface includes the mixed-type ($m'$-type) bands, a uniform pairing state emerges and competes with two other ordered states: a nematic state characterized by broken $C_3$ symmetry, and a $C_3$-chiral state with noncollinear phases among the three bonds. 

As further-neighbor hoppings bring the $m'$- and $p$-type bands close in energy near the $M$ point, the uniform state gives way to an additional chiral state driven by interband pairing. The origin of this state can be understood from the pairing matrix at the $M$ point, where inter-sublattice pairing is transformed into interband pairing in the band basis. Maximizing the interband pairing energy requires a relative phase difference of $\pi$ between two bond pairings, a condition that cannot be simultaneously satisfied for all three bonds. The resulting phase frustration is relieved by a $2\pi/3$ phase winding among the three bond pairings, stabilizing a chiral interband pairing state that spontaneously breaks time-reversal symmetry.

Since these features arise from the unique geometry and symmetry-enforced sublattice structure of the Kagome lattice in the strongly correlated regime, our results establish correlated Kagome systems as a promising platform for PDW, chiral, and nematic superconductivity. Going beyond the MF approximation may reveal additional correlated phases and competing orders, particularly given the weak phase locking of the PDW and recent evidence for chiral PDW states in large-$U$ models~\cite{YaoM2025,GenuinePDW}. Subleading interactions, including correlated hopping and further-neighbor exchange terms, may further enrich the phase diagram. From a materials perspective, identifying candidate materials that realize the distinct van Hove regimes discussed here remains an important open challenge. Once such systems are established, scanning tunneling microscopy and Kerr-effect measurements can provide direct probes of the nematic and chiral pairing states, respectively.

{\it Acknowledgments} - 
This work is supported by the Natural Sciences and Engineering Research Council of Canada (NSERC) Discovery Grant 2022-04601 and NSERC CREATE/575280-2023. H. Y. K. acknowledges support from the Canada Research Chairs Program CRC-2019-00147, and the hospitality of the Korea Institute for Advanced Study, where part of this work was carried out. This research was enabled in part by support provided by Calcul Québec and the Digital Research Alliance of Canada.

\bibliography{biblio}

\end{document}